\documentclass[%
 reprint,
 aps, prl
]{revtex4-2}

\usepackage{bm,color}
\usepackage{graphicx}
\usepackage{amsmath, amssymb}
\usepackage{color}
\usepackage{braket}
\usepackage{mathdots}
\usepackage{comment}

\begin{document}
\title {Nonperturbative effects in second harmonic generation}
\author{Keisuke Kitayama}
\affiliation{Department of Physics, Waseda University, Shinjuku, Tokyo 169-8555, Japan}
\author{Masao Ogata}
\affiliation{Department of Physics, University of Tokyo, Hongo, Bunkyo-ku, Tokyo 113-0033, Japan}
\affiliation{Research Institute for Energy Efficient Technologies, National Institute of Advanced Industrial Science and Technology (AIST), Umezono, Tsukuba, Ibaraki 305-8568, Japan}
\begin{abstract}
Second-harmonic generation (SHG) is a quintessential probe of inversion symmetry breaking in condensed matter. While perturbative $\chi^{(2)}$ processes are well-documented, the nonperturbative regime under intense driving remains largely unexplored. In this Letter, we develop a nonperturbative Floquet-Keldysh theory to describe SHG in two-band systems. Our analysis reveals the emergence of two distinct types of nonperturbative saturation: a transition from the conventional $E^2$ scaling to a linear $E$ dependence, and a stronger saturation regime where the SHG response becomes independent of the field amplitude. These behaviors are analytically shown to be governed by one-photon and two-photon resonance processes, respectively. By applying our formalism to a tight-binding model of monolayer GeS, we demonstrate that these specific scaling behaviors are observable in realistic materials and are fully consistent with large-scale numerical Floquet-matrix calculations.
\end{abstract}
\maketitle

\textit{Introduction.}---Second-harmonic generation (SHG), a fundamental nonlinear optical process in which an incident light field with frequency $\omega$ is converted to a response with frequency $2\omega$, has been a cornerstone of condensed matter physics and optics for decades~\cite{PhysRevLett.7.118, PhysRev.128.606, PhysRevLett.8.404, PhysRevLett.16.986, butet2015optical}. Since its discovery, SHG has been extensively investigated from both fundamental and applied perspectives. On the applied side, it serves as a powerful and versatile tool for probing the structural properties of materials, particularly for detecting the breaking of inversion symmetry in crystals and at interfaces~\cite{Lupke:95, doi:10.1126/sciadv.abe8691}. This sensitivity has led to its widespread use in materials science, surface science, and advanced imaging techniques such as SHG microscopy in biological systems~\cite{shen1989surface}. From a theoretical standpoint, SHG provides deep insights into the electronic band structure, bonding properties, and more recently, the geometric phases of quantum states in solids~\cite{PhysRevB.48.11705, PhysRevB.57.3905, doi:10.1126/sciadv.1501524}, making it an indispensable spectroscopic method.

Theoretically, SHG in the weak-field regime is well understood within the framework of perturbation theory~\cite{PhysRevB.48.11705}. The response is described by an expansion of the material's polarization in powers of the incident electric field, $E(\omega)$. The leading term responsible for SHG is the second-order polarization, $P(2\omega) = \chi^{(2)}E(\omega)E(\omega)$, where $\chi^{(2)}$ is the second-order nonlinear susceptibility tensor. This perturbative approach, which predicts an SHG intensity proportional to the square of the incident light intensity ($I_{2\omega} \propto I_{\omega}^2$), has been remarkably successful in explaining a vast range of experimental observations under moderate light intensities~\cite{PhysRevLett.8.404}. However, this framework is inherently limited to the weak-field regime. The behavior of SHG under intense laser fields, where nonperturbative effects beyond the standard $\chi^{(2)}$ descriptions become dominant, has remained largely unexplored despite its potential for high-speed optoelectronics.

The growing importance of such nonperturbative phenomena has recently been highlighted in the study of other second-order nonlinear responses, most notably the shift current~\cite{doi:10.1126/sciadv.abe8691, PhysRevB.23.5590, PhysRevLett.109.116601, belinicher1980photogalvanic}. The shift current is a DC photocurrent generated in noncentrosymmetric materials, which, like SHG, is formally a $\chi^{(2)}$ process related to the geometric phase of electronic wavefunctions. 
~\cite{doi:10.1126/sciadv.abe8691, doi:10.1126/sciadv.1501524, PhysRevB.110.045127, doi:10.1126/science.aaz9146}. The current deviates from the expected perturbative scaling ($J \propto E^2$) and transitions to a linear dependence on the field amplitude ($J \propto E$). These discoveries underscore that strong light-matter coupling can dramatically alter the nature of nonlinear optical responses, suggesting that similar nonperturbative effects will be latent in SHG as well.

In this Letter, we theoretically investigate the nonperturbative regime of SHG in solids by employing the nonperturbative Floquet-Keldysh theory, a powerful formalism for analyzing quantum systems under periodic driving~\cite{annurev:/content/journals/10.1146/annurev-conmatphys-031218-013423, RevModPhys.86.779, kitayama2020predicted, kitayama2021floquet, kitayama2022predicted}. Our analysis reveals that the SHG response undergoes a significant modification as the driving field intensity increases, leading to the emergence of two distinct saturation regimes. When the one-photon resonance dominates, the SHG signal deviates from the conventional $E^2$ scaling and transitions to a linear $E$ dependence. On the other hand, when the two-photon resonance dominates, we uncover stronger saturation mechanism, the transition from the parabolic to a constant saturation, rendering the SHG response independent of the field amplitude which is unexpected in the case of shift currents. By applying our theory to monolayer GeS, a gapped Dirac material known for its giant shift-current responsivity~\cite{cook2017design}, we demonstrate the experimental feasibility of observing these two types of nonperturbative saturation. These findings provide a fundamental understanding of SHG in the previously uncharted nonperturbative territory and open new avenues for controlling optical nonlinearities using intense light fields.

\begin{figure}[htb]
	\begin{center}
		\includegraphics[scale=1.0]{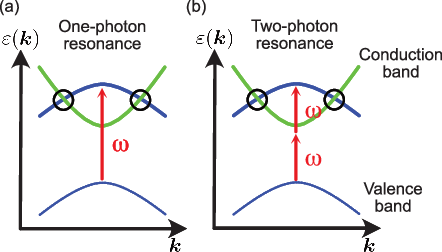}
		\caption{Schematics of (a) one-photon resonance process and (b) two-photon resonance process.}
		\label{Fig01}
	\end{center}
\end{figure}

\textit{Floquet-Keldysh Formalism.}---Here, we derive the equations that describe the nonperturbative effects of second-harmonic generation. We focus on a two-band system under irradiation with monochromatic light, which corresponds to a vector potential $\bm{A}(t) = i\bm{E} e^{i\Omega t}/\Omega - i\bm{E}^* e^{-i\Omega t}/\Omega$. Here, $\bm{E}$ denotes a vector with complex entries that corresponds to the light amplitude, and $\Omega$ is the frequency of light. Note that the complex amplitude allows us to deal with light of an arbitrary degree of polarization. However, in the present paper, we only assume irradiation with linearly polarized light, so all components can be chosen to be real. 

We assume that the time-dependent Hamiltonian is given by the Peierls substitution $H(t) = H_0(\bm{k} + e\bm{A}(t)/\hbar)$, where $H_0(\bm{k})$ is the Bloch Hamiltonian without the drive. To explicitly map this time-periodic Hamiltonian onto a static Floquet Hamiltonian $H_{\rm F}$, we first expand $H(t)$ in powers of the vector potential up to the second order:
\begin{equation}
    H(t) \simeq H_0(\bm{k}) + \frac{e}{\hbar} A_\alpha(t) \frac{\partial H_0}{\partial k_\alpha} + \frac{e^2}{2\hbar^2} A_\alpha(t) A_\beta(t) \frac{\partial^2 H_0}{\partial k_\alpha \partial k_\beta},
\end{equation}
where the summation over repeated spatial indices $\alpha, \beta$ is implied. This expansion allows us to extract the Fourier components $H_n = \frac{1}{T} \int_0^T H(t) e^{in\Omega t} dt$. The non-zero relevant components are given by:
\begin{align}
    H_1 &= i\frac{e}{\hbar\Omega} \bm{E} \cdot \frac{\partial H_0}{\partial \bm{k}}, \\
    H_{-1} &= -i\frac{e}{\hbar\Omega} \bm{E}^* \cdot \frac{\partial H_0}{\partial \bm{k}}, \\
    H_2 &= -\frac{e^2}{2\hbar^2\Omega^2} E^\alpha E^\beta \frac{\partial^2 H_0}{\partial k_\alpha \partial k_\beta},
\end{align}
along with $H_{-2} = H_2^*$ (for real $E^\alpha, E^\beta$, $H_{-2} = H_2$). The static Floquet Hamiltonian in the extended photon-dressed state space is defined as $[H_{\rm F}]_{n,m} = H_{n-m} - m\hbar\Omega\delta_{n,m}$. 

In the following, we consider the situation where the system is coupled to a fermionic bath at temperature $T_{\rm bath}$ with a dissipation coefficient $\Gamma$. The expression for the time-dependent current is given by
\begin{equation}
J^\mu(t) = -i \sum_{m,n}\int \frac{d^d k}{(2\pi)^d} \int\frac{d\omega}{2\pi}\mathrm{Tr} [v^\mu_{n,m} G^<_{mn}(\omega)] e^{-i(n-m)\Omega t}, \label{eq::J(t)}
\end{equation}
where $v^\mu_{n,m}$ are the matrix elements of the current operator $\bm{v}(t) = \partial H(t) / \hbar\partial \bm{k}$, and $G^<$ denotes the lesser Green's function in the Keldysh formalism. In the case of SHG, we extract the frequency $2\Omega$ component from Eq.~(\ref{eq::J(t)}), corresponding to $n - m = 2$:
\begin{equation}
J^\mu(2\Omega) = -i \sum_{m}\int \frac{d^d k}{(2\pi)^d} \int\frac{d\omega}{2\pi}\mathrm{Tr} [\tilde{v}^\mu_{m+2,m} G^<_{m,m}(\omega)], \label{eq::J(2omega)}
\end{equation}
where $\tilde{\bm{v}} = \partial \mathcal{H}_{\rm F}/\hbar\partial \bm{k}$.

This SHG response stems from two distinct physical processes, distinguished by how the system traverses the Floquet states: the one-photon resonance and the two-photon resonance processes [Fig.~\ref{Fig01}]~\cite{PhysRevB.94.035117}.

To evaluate the one-photon contribution, we isolate the resonance condition $\varepsilon_1 + \hbar\Omega \approx \varepsilon_2$, where the subscripts 1 and 2 stand for the valence and conduction bands, respectively. By invoking the rotating-wave approximation (RWA), we can neglect far-off-resonant blocks in the Floquet Hamiltonian and safely truncate the full Floquet matrix into a $2 \times 2$ effective Hamiltonian spanned by the valence band dressed with one photon ($m=1$) and the conduction band with zero photons ($m=0$). The coupling between these states is mediated entirely by $H_1$ and $H_{-1}$. The truncated Hamiltonian reads:
\begin{equation}
H_{\rm F}^{\text{1ph}} = \left(
	\begin{array}{cc}
		 \varepsilon_1 + \hbar\Omega & -i\frac{e\bm{E}^*}{\hbar\Omega}\cdot \left( \frac{\partial H_0(\bm{k})}{\partial \bm{k}} \right)_{12} \\
		 i\frac{e\bm{E}}{\hbar\Omega}\cdot \left( \frac{\partial H_0(\bm{k})}{\partial \bm{k}} \right)_{21} & \varepsilon_2
	\end{array}
	\right).
	\label{eq::FloquetHam1}
\end{equation}
Solving the Dyson equation for this $2 \times 2$ Keldysh space yields the lesser Green's function, from which we obtain the one-photon contribution:
\begin{align}
J_{\rm 1ph}^\mu &= -\frac{2\pi e^3 |\bm{E}|^2}{\hbar^2}\int \frac{d^d k}{(2\pi)^d} (f_1 - f_2)S^\mu_{\bm{k}}\left|\left(\frac{\partial H_0(\bm{k})}{\partial \bm{k}}\right)_{12}\right|^2 \nonumber \\
& \times \frac{\frac{\Gamma}{2}}{ \sqrt{\left|\frac{e\bm{E}}{\hbar\Omega}\cdot \left( \frac{\partial H_0(\bm{k})}{\partial \bm{k}} \right)_{12}\right|^2 + \frac{\Gamma^2}{4}}}\delta(\varepsilon_1-\varepsilon_2+\hbar\Omega).
\label{eq::J1ph}
\end{align}
Here, $|\bm{E}|^2 = \bm{E}^*\cdot \bm{E}$ and $f_{1,2} = 1/\{1 + \exp[(\varepsilon_{1,2}-\mu)/k_BT_{\rm bath}] \}$ is the Fermi-Dirac distribution. The shift vector is defined as $S^\mu_{\bm{k}} = \partial \varphi_{1,2} / \partial k_\mu + A_1^\mu(\bm{k}) - A_2^\mu(\bm{k})$, where $\varphi_{1,2}$ is the phase of the interband velocity matrix element and $A_{n}^\mu(\bm{k})$ is the intraband Berry connection. This saturation factor shares the same mathematical structure as that derived for the nonperturbative shift current~\cite{doi:10.1126/sciadv.1501524}.

Conversely, the two-photon contribution arises from the direct $2\hbar\Omega$ resonance condition ($\varepsilon_1 + 2\hbar\Omega \approx \varepsilon_2$). To capture this, we apply the RWA by selecting a different $2 \times 2$ subspace from the Floquet Hamiltonian: the valence band dressed with two photons ($m=2$) and the conduction band with zero photons ($m=0$). The direct coupling between these sectors is provided by the $H_2$ component derived from the diamagnetic ($\bm{A}^2$) term. The corresponding truncated Floquet Hamiltonian is:
\begin{equation}
H_{\rm F}^{\text{2ph}} = \left(
	\begin{array}{cc}
		 \varepsilon_1 + 2\hbar\Omega & -\frac{1}{2}\frac{e^2(E^\alpha E^\beta)^*}{\hbar^2\Omega^2} \left( \frac{\partial^2 H_0(\bm{k})}{\partial k_\alpha \partial k_\beta} \right)_{12} \\
		  -\frac{1}{2}\frac{e^2E^\alpha E^\beta}{\hbar^2\Omega^2} \left( \frac{\partial^2 H_0(\bm{k})}{\partial k_\alpha \partial k_\beta} \right)_{21} & \varepsilon_2
	\end{array}
	\right).
	\label{eq::FloquetHam2}
\end{equation}
Employing the same Keldysh formalism applied to this subspace, we calculate the lesser Green's function to find the two-photon contribution:
\begin{align}
J_{\rm 2ph}^\mu &= \frac{\pi e^3 |\bm{E}|^2}{\hbar^2}\int \frac{d^d k}{(2\pi)^d} (f_1 - f_2)S^\mu_{\bm{k}}\left|\left(\frac{\partial H_0(\bm{k})}{\partial \bm{k}}\right)_{12}\right|^2 \nonumber \\
& \times \frac{\Gamma}{ \sqrt{ \left|\frac{e^2 E^\alpha E^\beta}{\hbar^2\Omega^2} \left( \frac{\partial^2 H_0(\bm{k})}{\partial k_\alpha \partial k_\beta} \right)_{12} \right|^2 + \Gamma^2} }\delta(\varepsilon_1-\varepsilon_2+2\hbar\Omega),
\label{eq::J2ph}
\end{align}
where the summation of the indices $\alpha$ and $\beta$ are abbreviated in the denominator of the saturation factor. By combining the one-photon and two-photon contributions, we obtain a comprehensive expression for the nonperturbative SHG current, $J^{\mu}(2\Omega) = J_{\rm 1ph}^\mu + J_{\rm 2ph}^\mu$. 
This analytical result reveals two distinct saturation behaviors governed by the field-dependent denominators. 
In the frequency regime where the one-photon resonance dominates, the SHG response undergoes a transition from the conventional perturbative scaling ($J \propto E^2$) to a linear dependence on the field amplitude ($J \propto E$). 
On the other hand, when the two-photon term $J_{\rm 2ph}$ becomes significant, even more pronounced saturation mechanism where the SHG current becomes independent of the field (constant saturation). 
Notably, this plateauing effect represents a novel nonperturbative feature that has not been previously reported in nonlinear optical studies, distinguishing the two-photon resonance as a regime of stronger suppression compared to the one-photon counterpart.

We emphasize that our derived equations are inherently consistent with perturbation theory. 
In the limit of low light intensities ($E \to 0$), the field-dependent terms in the denominators of Eqs.~(\ref{eq::J1ph}) and (\ref{eq::J2ph}) become negligible compared to the dissipation rate $\Gamma$. 
In this regime, the current recovers the standard parabolic scaling with respect to the electric field, in perfect agreement with the traditional $\chi^{(2)}$ description. 
The transition to saturation thus highlights the critical role of strong light-matter coupling that lies beyond the reach of standard perturbative expansions.

\textit{Application to Monolayer GeS.}---To demonstrate the experimental relevance of our theory, we apply the formalism to monolayer GeS, a member of the orthorhombic group-IV monochalcogenides. GeS is characterized as a gapped Dirac material with broken inversion symmetry ($C_{2v}$ point group), which has attracted significant attention for its large shift-current response in photovoltaic applications~\cite{cook2017design}. The strong light-matter coupling inherent in its electronic structure makes it an ideal platform for exploring nonperturbative SHG.

We describe the electronic states of GeS using a two-band tight-binding model on a rectangular lattice, where the hopping parameters are determined from first-principles calculations to accurately capture the band-edge physics near the $\Gamma$ point. The explicit form of Hamiltonian is given in Ref.~\cite{cook2017design}. We evaluate the SHG response using two distinct approaches: (Method 1) the analytical expressions for $J_{\rm 1ph}$ and $J_{\rm 2ph}$ derived from the truncated $2 \times 2$ Floquet space [Eqs.~(\ref{eq::J1ph}) and (\ref{eq::J2ph})], and (Method 2) a full numerical evaluation of the Floquet-Keldysh formula [Eq.~(\ref{eq::J(2omega)})] using a $22 \times 22$ truncated Floquet matrix (corresponding to photon indices $m \in [-5, 5]$ for the two-band system). While Method 1 focuses on specific resonance channels, Method 2 accounts for high-order multiphoton processes and power broadening without further approximation.

\begin{figure}[t]
	\begin{center}
		\includegraphics[scale = 0.9]{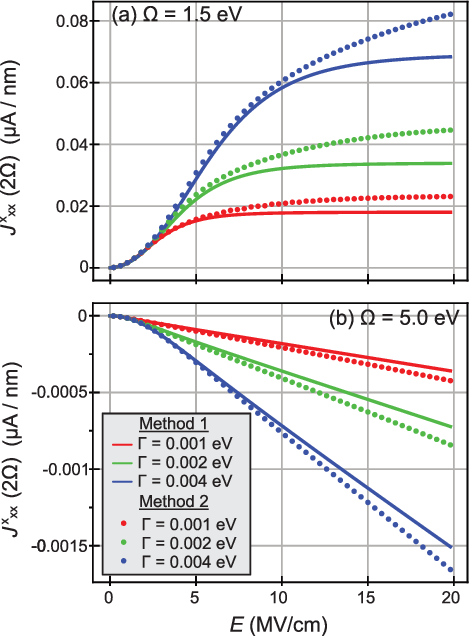}
		\caption{Calculated SHG current in GeS as a function of the incident light amplitude $E$. (a) Sub-gap driving at $\hbar\Omega = 1.5$~eV. (b) Above-gap driving at $\hbar\Omega = 5.0$~eV. Lines and dots represent Method 1 (analytical) and Method 2 (numerical), respectively.}
		\label{Fig02}
	\end{center}
\end{figure}

Figure~\ref{Fig02} displays the SHG current $J(2\Omega)$ for two representative driving frequencies. For sub-gap excitation at $\hbar\Omega = 1.5$~eV (upper panel), the one-photon resonance is energetically forbidden ($\Omega < E_g$), leaving the two-photon resonance as the primary excitation channel. In this regime, we observe that the SHG signal initially follows the perturbative $E^2$ scaling but saturates to a constant value at high light intensities. This plateau is a direct consequence of the field-dependent denominator in $J_{\rm 2ph}$ (Eq.~(\ref{eq::J2ph})), where the intensity-induced broadening compensates for the increased driving strength.

In stark contrast, for $\hbar\Omega = 5.0$~eV (lower panel), the one-photon resonance becomes the dominant contribution. Here, the SHG response exhibits a crossover from $E^2$ to a clear linear dependence on the field amplitude ($J \propto E$) as the intensity increases. This linear scaling persists even when considering high-order photon transitions in the $22 \times 22$ Floquet calculation (Method 2), confirming that the one-photon saturation mechanism described by Eq.~(\ref{eq::J1ph}) is robust. 

The excellent agreement between the analytical lines and numerical dots across both frequency regimes validates our RWA-based truncation approach. The emergence of these two distinct saturation behaviors---constant plateauing versus linear scaling---within a single material underscores the rich nonperturbative landscape of GeS. Given its stability and the accessibility of its nonlinear responses, we propose monolayer GeS as a promising candidate for experimental verification of these saturation regimes in second-order optical processes.

\textit{Discussion.}---The crossover from perturbative to nonperturbative regimes is physically governed by the competition between the Rabi frequency $\Omega_R \sim eE|\bm{v}_{12}|/\hbar\Omega$ and the relaxation rate $\Gamma$. 
When $\Omega_R \gg \Gamma$, the carrier populations and coherences reach a steady state that no longer follows the $E^2$ scaling, a phenomenon analogous to the power broadening in quantum optics. 
In the case of GeS, given the large interband dipole moment $d \approx 2.5$~\AA\ near the $\Gamma$ point~\cite{cook2017design}, the saturation threshold occurs at electric field strengths of approximately $10^7$--$10^8$~V/m. 
Such field strengths are readily accessible using modern mid-infrared or terahertz ultrafast laser sources without exceeding the optical damage threshold of monolayer chalcogenides.

Furthermore, it is remarkable that the analytical $2 \times 2$ model (Method 1) remains in excellent agreement with the full $22 \times 22$ Floquet calculation (Method 2) even at extreme fields. 
This suggests that for SHG, higher-order photon-dressing processes primarily contribute to a renormalization of the effective bandgap (dynamic Stark shift) rather than introducing entirely new resonance channels, provided the driving frequency is not in a high-order multiphoton resonance with deeper valence bands. 
Our findings thus provide a simplified yet robust framework for predicting nonlinear optical responses in the strong-field limit.

An alternative perspective on our results is that strong-field SHG can serve as a direct probe of the underlying Floquet resonance structure in driven crystals. The two distinct saturation behaviors identified here are not merely different degrees of suppression, but qualitative fingerprints of the dominant photon-dressed excitation pathway. A crossover from the perturbative quadratic response to an approximately linear behavior indicates that the SHG process is primarily governed by a one-photon resonant channel, whereas the emergence of a plateau signals that a two-photon channel has become dominant. From this viewpoint, the field dependence of SHG contains information that is absent in the weak-field susceptibility alone, because it reveals how periodic driving reorganizes the available interband processes in the nonequilibrium steady state. The close agreement between the reduced Floquet description and the full numerical calculation further suggests that these scaling forms are robust and experimentally useful as diagnostics of the resonance mechanism in realistic materials.

\textit{Conclusions.}---In summary, we have developed a nonperturbative theory of SHG using the Floquet-Keldysh formalism. 
We identified two saturation regimes and demonstrated their presence in monolayer GeS. 
Our results extend the understanding of nonlinear optics into the nonperturbative territory and suggest that intense light can be used to dynamically tune and suppress second-order optical responses in topological materials.

\textit{Acknowledgment.}---This work was supported by JSPS KAKENHI (No. 21J20856, No. 23K03274, and No. 25K17352).
\bibliography{crossresp}
\end{document}